# Transport properties and anisotropy in rare earth doped $CaFe_2As_2$ single crystals with $T_c$ above 40 K

Yanpeng Qi[*], Zhaoshun Gao[*], Lei Wang, Dongliang Wang, Xianping Zhang, Chao Yao, Chunlei Wang, Chengduo Wang, Yanwei Ma[1]

Key Laboratory of Applied Superconductivity, Institute of Electrical Engineering, Chinese Academy of Sciences, P. O. Box 2703, Beijing 100190, China

**Abstract:**

In this paper we report the superconductivity above 40 K in the electron doping single crystal $Ca_{1-x}Re_xFe_2As_2$ (Re = La, Ce, Pr). The x-ray diffraction patterns indicate high crystalline quality and c-axis orientation. the resistivity anomaly in the parent compound $CaFe_2As_2$ is completely suppressed by partial replacement of Ca by rare earth and a superconducting transition reaches as high as 43 K, which is higher than the value in electron doping FeAs-122 compounds by substituting Fe ions with transition metal, even surpasses the highest values observed in hole doping systems with a transition temperature up to 38 K. The upper critical field has been determined with the magnetic field along *ab*-plane and *c*-axis, yielding the anisotropy of 2~3. Hall-effect measurements indicate that the conduction in this material is dominated by electron like charge carriers. Our results explicitly demonstrate the feasibility of inducing superconductivity in Ca122 compounds via electron doping using aliovalent rare earth substitution into the alkaline earth site, which should add more ingredients to the underlying physics of the iron-based superconductors.

---

[*] Joint first authors. Both authors contributed equally to this work.

[1] Author to whom correspondence should be addressed; E-mail: ywma@mail.iee.ac.cn

# 1. Introduction

The discovery of new high-temperature superconductors based on FeAs has led to a new "gold rush" in high-$T_c$ superconductivity. Since the discovery of superconductivity at $T_c$ = 26 K in the fluorine doped LaFeAsO [1], the superconducting transition temperature $T_c$ can be raised up to 55 K by replacing La with other rare earth elements [2]. More and more new iron-based compounds with different structures have been found, thus a whole class of iron-based high-temperature superconductors has been established. Until now, the new iron based superconductors can be divided into two classes. The first class are iron pnictide materials, including ReFeAsO with ZrCuSiAs-type structure [1-6], BaFe$_2$As$_2$ with ThCr$_2$Si$_2$-type structure [7-10], LiFeAs with PbFCl-type structure [11,12]. They contain two dimensional FeAs tetrahedron layers and atomic sheets or oxide blocks along the c axis, which act as charge reservoirs. Another class are iron chalcogenides FeCh (Ch = chalcogens): binary FeCh-11 type FeSe compounds [13] and recently discovered new A$_x$Fe$_2$Se$_2$ system by introducing charge reservoir layers between FeCh sheets [14]. All compounds are layered materials with a common layer of an FeAs(Se) in between different spacing layers. In these compounds superconductivity appears close to the disappearance of an antiferromagnetic (AFM) state upon doping or application of pressure, suggesting an unconventional superconducting state in analogy to those observed in high-$T_c$ or heavy-fermion superconductors.

It is well known that the existence of the symmetry between electron- and hole-doping has been demonstrated in cuprate high temperature superconductors with respect to the induction of superconductivity in their respective parent compounds [15]. The first iron based superconductors LaFeAsO$_{1-x}$F$_x$ by substituting part of the oxygen with fluorine has been proved to be bearing electron type carriers [1], quickly superconductivity could also be achieved by substituting La$^{3+}$ with Sr$^{2+}$, that is thought hole doping [5]. The highest transition temperature is about 25 K either in electron doping system or hole doping case. On the other hand, in FeAs122 system, hole doped Ba$_{1-x}$K$_x$Fe$_2$As$_2$ have been reported to realize superconductivity at 38 K[7], however, the superconductivity can be realized at lower temperature around 22 K by substituting Fe ions with other transition metals, which can introduce more electrons into BaFe$_2$As$_2$ [16]. Thus the possibility to raise the transition temperature by electron doping in FeAs122 system is very attractive. In fact, the attempt has been made by substituting Ba$^{2+}$ with rare earths elements and no trace of superconductivity could be observed in Ba$_{1-x}$La$_x$Fe$_2$As$_2$ samples synthesized by the conventional sold state reaction method [17]. Very recently, Saha S. R. et al. have employed rare earth substitution into CaFe$_2$As$_2$ [18]. It is very surprise that superconductivity up to 45 K induced by replacing the Ca sites with rare earth element in CaFe$_2$As$_2$ compounds. In present work, we report the fabrication and characterization of the electron-doped Ca$_{1-x}$Re$_x$Fe$_2$As$_2$ (Re = La, Ce, Pr) single crystals. X-ray diffraction (XRD), scanning electron microscope energy dispersive x-ray (SEM-EDX), resistivity, and dc magnetic susceptibility as well as Hall-effect measurements have been performed for the Ca$_{1-x}$Re$_x$Fe$_2$As$_2$ samples. Superconductivity up to 43 K was found in Ca$_{1-x}$Re$_x$Fe$_2$As$_2$, which exceeds the theoretical maximum predicted by the Bardeen-Cooper-Schrieffer theory. Our results demonstrate that superconductivity can be realized by replacing the Ca sites with rare earth electrons, which would add more ingredients to the underlying physics of the iron-based superconductors.

**2. Experimental**

The single crystals of Ca$_{1-x}$Re$_x$Fe$_2$As$_2$ (Re = La, Ce, Pr) were grown by the FeAs self-flux method. The details of fabrication process are described elsewhere [19]. The FeAs precursor was synthesized by the reaction of Fe powder and As chips at 500 °C for 10 h and then 700 °C for 20 h in a sealed Niobium tube. Then Re-pieces and Ca-pieces were mixed with FeAs according to the ratio of (Re+Ca)/FeAs = 1/4 and placed in an alumina crucible inside a pure iron tube sealed under reduced Ar atmosphere. The assembly was put into tube furnace and heated to 1200 °C slowly and held there for 5 hours, and then was cooled to 1030 °C with a rate of 3 ~ 6 °C/h, finally it was furnace-cooled to room temperature by turning off the power. It is noted that all the weighing, mixing and encasing procedures were performed in a glove box in which high pure argon atmosphere is filled. Single crystals with the flat shiny surface up to 5 mm x 5mm size were obtained by cleaving the as-grown single crystals.

Single crystals were characterized by x-ray diffraction (XRD, *Mac-Science* MXP18A-HF equipment) using Cu $K_{\alpha}$ radiation from 10° to 80°. The crystal surface morphology and composition were examined by scanning electron microscopy (SEM, Hitachi S-4800) and the energy dispersive X-ray (EDX, Oxford-6566, installed in the S-4200 apparatus) analysis. DC magnetization measurement was carried out on a Quantum Design physical property measurement system (VSM, PPMS). The resistance and Hall effect measurements were done using six-probe technique on the physical property measurement system (Quantum Design) with the magnetic field up to 9 T.

## 3. Results and discussion

Figure 1 show the single crystal x-ray diffraction (XRD) patterns for the $CaFe_2As_2$ and $Ca_{1-x}Re_xFe_2As_2$ (Re = La, Ce, Pr). Only (*00l*) peaks were observed in the single crystal XRD patterns, indicating that the crystallographic *c* axis is perpendicular to the plane of the single crystal. It is found that the position of (00*l*) peaks of Re doped sample shift to higher angles, suggesting a successful chemical substitution and decrease in *c*-axis lattice. The change of *c*-axis lattice parameters of in $Ca_{1-x}Re_xFe_2As_2$ compounds is contrary to the case of Na-doping in $Ca_{1-x}Na_xFe_2As_2$ compounds [9].

Microstructure observations were performed using scanning electron microscope (SEM/EDX). The as-grown crystal shows a very flat surface morphology and the layered structure. The typical EDX spectrum is shown in Figure 2. From the EDX analysis we found only the right components, i.e., Ca, Re, Fe, and As, without any other foreign elements. The actual compositions of the single crystals are close to the nominal ones. It should be noted that the same was found for other nominal compositions of samples.

In Fig. 3, we present the temperature dependence of the normalized resistance for single crystals of $Ca_{1-x}Re_xFe_2As_2$ (Re =La, Ce, Pr). The parent $CaFe_2As_2$ compound exhibits a resistivity anomaly at about 160 K, which is ascribed to the magnetic/structural phase transition [9, 20-21]. Similar to the case of substituting Ca ions with Na or Fe ions with other transition metals, the abnormal behavior in the resistivity is clearly suppressed and resistivity drop is converted to a broader uprising in the lower temperature [9, 22]. Take the case of La-doped $Ca_{1-x}Re_xFe_2As_2$ compounds for example, the resistivity anomaly shifted to a lower temperature at around 100 K for the sample x = 0.1. Interestingly, ρ decreases quickly below 10 K, suggesting a superconducting transition, however, no zero resistivity was observed down to the lowest temperature, probably because of the much broader transition width in this sample. On the doping 20% La, the anomaly is completely suppressed and a superconducting transition reaches as high as 42.7 K. the superconducting transition is rather sharp with transition width about 3 K (90% - 10% of normal state resistivity). Similar phenomena were observed in $Ca_{1-x}Re_xFe_2As_2$ and $Ca_{1-x}Re_xFe_2As_2$ samples (Fig. 3 b, c) and the transition temperature reaches the value of 43 K for $Ca_{1-x}Ce_xFe_2As_2$ and 43.5 K for $Ca_{1-x}Pr_xFe_2As_2$, respectively. It should be noted that there exist two transition steps on the resistivity curve measured under zero field, one at about 43 K and another at about 20 K, which is consistent with previous reports [23]. Our data demonstrate that superconductivity in $CaFe_2As_2$ could achieved by substituting Ca ions with other rare earth elements.

In order to further confirm the superconductivity of our samples, DC magnetic susceptibility measurement was also performed. Fig. 4 shows the temperature dependence of the DC magnetization for the $Ca_{1-x}Re_xFe_2As_2$ (Re =La, Ce, Pr) samples. Although the transitions are still broad, an enlarged view shown in the inset allow us to determine the onset magnetic transition point. It is clear that a clear diamagnetic signal appears around 40 K in the samples, which correspond to the middle transition temperature of the resistivity data. It should be noted that the magnetic transition curves are still not perfectly sharp, which leaves more room for improving the sample quality in the future work. But this does not give any doubt about the superconducting transition temperatures determined here. It is very interesting that superconductivity could be induced in Ca122 system by substituting Ca ions with other rare earth elements, which is similar to the case of rare earth electron-doping in the fluorine-based materials (Ca,Re)FeAsF [24]. More surprisingly, rare earth substitution into CaFe$_2$As$_2$ induces superconductivity with transition temperature as high as 43 K. These values not only have exceeded previous ones found in any FeAs-122 system, including both the commonly alkali metals elements doping and transition metal doping at Fe site, but also well exceeds the theoretical maximum predicted by the Bardeen-Cooper-Schrieffer theory.

The resistivity of Ca$_{0.8}$La$_{0.2}$Fe$_2$As$_2$ and Ca$_{0.8}$Ce$_{0.2}$Fe$_2$As$_2$ with the different magnitude of magnet field applied parallel and perpendicular to the ab-plane around T$_c$ are shown in Fig.5. The transition temperature of superconductivity is suppressed gradually and the transition is broadened with increasing the magnetic field, however, obvious difference for the effect of field along different direction on the superconductivity can observed. It is worth to note that the transition temperature is sensitive to magnetic field under the condition of *H//c*. The ρ-T curves shift obviously to lower temperature even in a field of 0.1 T, which is different from other iron based superconductors. We tried to estimate the upper critical field ($H_{c2}$) and irreversibility field *($H_{irr}$)*, using the 90% and 10% points on the resistive transition curves. The estimated upper critical field *($H_{c2}$)* and irreversible field *($H_{irr}$)* are plotted in Figs. 6a and 6b as a function of temperature, respectively. Within the weak-coupling BCS theory, the upper critical field at T=0 K can be determined by the Werthamer-Helfand-Hohenberg (WHH) equation [25] $H_{c2}(0) = 0.693 \times (dH_{c2}/dT) \times T_c$. Using $-dH_{c2}/dT|_{Tc}$ = 6.25 T/K for *H//ab* and $-dH^c_{c2}/dT|_{Tc}$ = 1.88 T/K for *H//c*, one can estimate the values of upper critical fields to be 185 T and 55.6 T with the field parallel and perpendicular to the ab-plane respectively for Ca$_{0.8}$La$_{0.2}$Fe$_2$As$_2$, respectively. For Ca$_{0.8}$Ce$_{0.2}$Fe$_2$As$_2$ crystal, we take $-dH_{c2}/dT|_{Tc}$ = 5.45 T/K for *H//ab* and $-dH^c_{c2}/dT|_{Tc}$ = 2.82 T/K for *H//c*. The H$_{c2}$(0) can be estimated to be 162.4 T and 84 T with the field parallel and perpendicular to the ab-plane, respectively. The anisotropy $H^{ab}_{c2}/H^c_{c2}$ is about 3.3 for Ca$_{0.8}$La$_{0.2}$Fe$_2$As$_2$ which is consistent with the value of 3.0~3.6 in AFe$_2$Se$_2$ [14, 26]. However the anisotropy is only 1.9 for a Ca$_{0.8}$Ce$_{0.2}$Fe$_2$As$_2$. Compared to the anisotropy in other FeAs-based superconductors, such as 5 in NdFeAsO$_{1-x}$F$_x$ [27], 1.7-1.86 in Ba$_{0.6}$K$_{0.4}$Fe$_2$As$_2$ [28], these values are all quite small compared to High-T$_c$ cuprates, which indicates an encouraging application perspective.

It is known that Hall effect measurement is a useful tool to investigate the information of charge carriers and the band structure. The Hall coefficient is almost independent of temperature for a conventional metal with Fermi liquid feature. However, this situation is changed for a multiband material [29] or a sample with non-Fermi liquid behavior, such as the cuprate superconductors [30]. To get more information about the conducting carrier of the Ca$_{1-x}$Re$_x$Fe$_2$As$_2$ (Re =La, Ce) single crystals, we also carried out the Hall-effect measurements. In the experiments Hall resistivity $\rho_{xy}$ was taken as $\rho_{xy}$ = [$\rho$(+H)- $\rho$(-H)]/2 at each point to eliminate the effect of the misaligned Hall electrodes. Moreover, $\rho_{xy}$ is negative at all the temperatures giving a positive Hall coefficient R$_H$ = $\rho_{xy}$ /H. The temperature dependence of R$_H$ is shown in Fig. 7. One can see that the Hall coefficient changes slightly at high temperatures but drops below 100 K. The negative Hall coefficient R$_H$ actually indicates that electron type charge carriers dominate the conduction in the present sample. In the iron based 1111 superconductors, the highest T$_c$ of 55 K occurs in the electron doped case. However, for 122 phase, the current record of 38 K is found in hole doped compound, which is obviously higher than the case of common electron doping via transition metal doping in Fe side. And now the superconductivity in the electron doped Ca122 system has exceeded the values of 40 K. Our results, combining with the previous reports [18, 23], demonstrate that T$_c$ could be enhanced by electron doping above the value of 38 K in FeAs-122 system.

The pristine CaFe$_2$As$_2$ is considered as a good springboard to investigating iron pnictide superconductivity. At ambient pressure CaFe$_2$As$_2$ undergoes a transition from a tetragonal to an orthorhombic phase [17, 20, 31-32]. When a hydrostatic pressure (p > 0.35 GPa) is applied to CaFe$_2$As$_2$, the orthorhombic phase transforms to a new collapsed tetragonal structure with the a-axis and c-axis lattice parameters undergoing a discontinuous jump. More surprisingly, the collapsed phase could be stabilized at ambient pressures by doping rare earth into CaFe$_2$As$_2$. It should be noted that different doping effects could be observed in $Ca_{1-x}Re_xFe_2As_2$ (Re =La, Ce, Pr). The neutron powder diffraction measurements of $CaFe_2As_2$ found that the substitution of Pr or Nd into $CaFe_2As_2$ could drive the collapse tetragonal transition; in contrast, the substitution of up to 28% La or 17% Ce does not drive the system toward any observable transition. The different ionic radii between doped Re and Ca ion possibly lead to various structure evolution, although similar superconductivity ~ 42 K could be observed in $Ca_{1-x}Re_xFe_2As_2$ (Re =La, Ce, Pr) compounds. It is clear that the interactions associated with structural, magnetic, and superconducting instabilities in the $CaFe_2As_2$ compounds are finely balanced and can be readily tuned through chemical substitution as well as pressure. The systemic research in those amazing compounds could add more ingredients to the underlying physics of the iron-based superconductors.

## 4. Conclusions

In summary, Single crystals of $Ca_{1-x}Re_xFe_2As_2$ (Re =La, Ce, Pr) are successfully synthesized with the superconducting transition temperatures $T_c$ above 40 K. The substitution of aliovalent rare earth into the Ca ions site could suppress the magnetic/structural phase transition in the parent $CaFe_2As_2$ compound and superconductivity reaches as high as 43 K, which exceeds previous values found in FeAs-122 compounds. We also determined the upper critical fields along ab-plane and c-axis. The anisotropy of the superconductor determined by the ratio of $H_{c2}^{ab}$ and $H_{c2}^{c}$ is estimated to 2~3. The Hall coefficient is negative, indicating the electrical transport behavior. Our results suggest that superconductivity can be realized in Ca122 compounds by replacing the alkaline earth sites with aliovalent rare earth elements.

**Acknowledgments**

The authors thank Profs. Haihu Wen, Liye Xiao and Liangzhen Lin for their help and useful discussions. This work is partially supported by National '973' Program (Grant No. 2011CBA00105) and the National Natural Science Foundation of China (Grant No. 51025726 and 51002150).

**Captions**

Figure 1 (Color online) X-ray diffraction patterns of the single crystal $Ca_{1-x}Re_xFe_2As_2$ (Re =La, Ce, Pr).

Figure 2 (Color online) The Energy dispersive X-ray microanalysis (EDX) spectrums of the samples $Ca_{1-x}Re_xFe_2As_2$ (Re =La, Ce, Pr).

Figure 3 (Color online) Temperature dependence of resistivity for the $Ca_{1-x}Re_xFe_2As_2$ (Re =La, Ce, Pr) samples measured in zero field. The data are normalized to $R_{300K}$.

Figure 4 (Color online) Temperature dependence of DC magnetization for the $Ca_{1-x}Re_xFe_2As_2$ (Re =La, Ce, Pr) samples. The measurements were done under a magnetic field of 20 Oe with zero-field-cooled and field-cooled modes.

Figure 5 (Color online) (a) and (b) show the temperature dependence of resistivity for $Ca_{0.8}La_{0.2}Fe_2As_2$ crystal with the magnetic field perpendicular and parallel to the c-axis respectively; (c) and (d) show the temperature dependence of resistivity for $Ca_{0.8}Ce_{0.2}Fe_2As_2$ crystal with the magnetic field perpendicular and parallel to the c-axis respectively;

Figure 6 (Color online) (a) and (b) show the temperature dependence of $H_{c2}$ and $H_{irr}$ for $Ca_{0.8}La_{0.2}Fe_2As_2$ and $Ca_{0.8}Ce_{0.2}Fe_2As_2$, respectively.

Figure 7 (Color online) Temperature dependence of the Hall coefficient $R_H$ for $Ca_{0.8}La_{0.2}Fe_2As_2$ and $Ca_{0.8}Ce_{0.2}Fe_2As_2$, respectively.

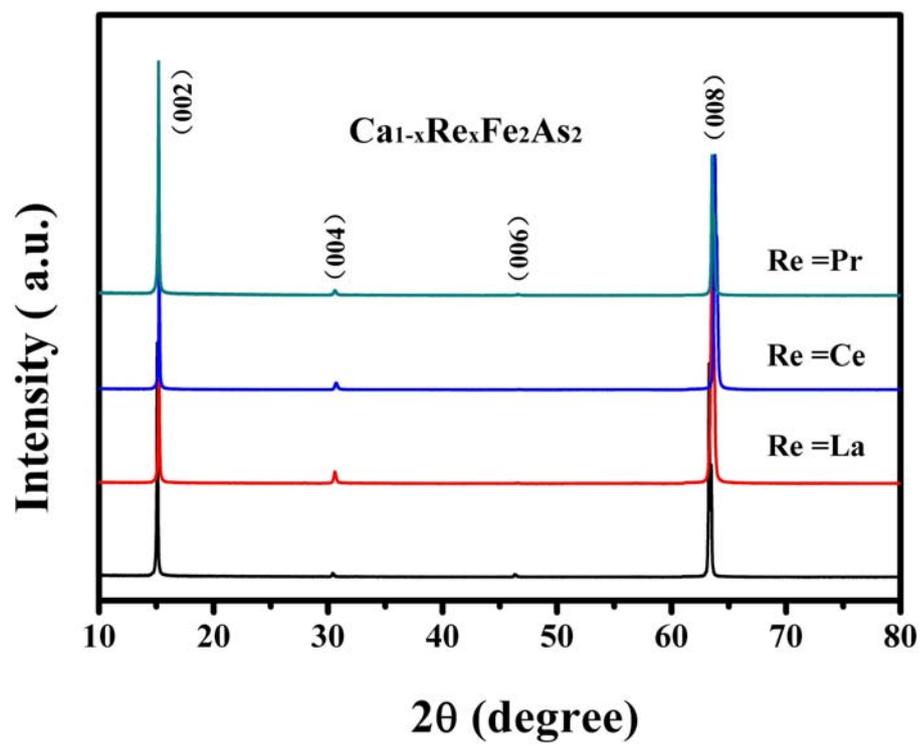

Fig.1 Qi et al.

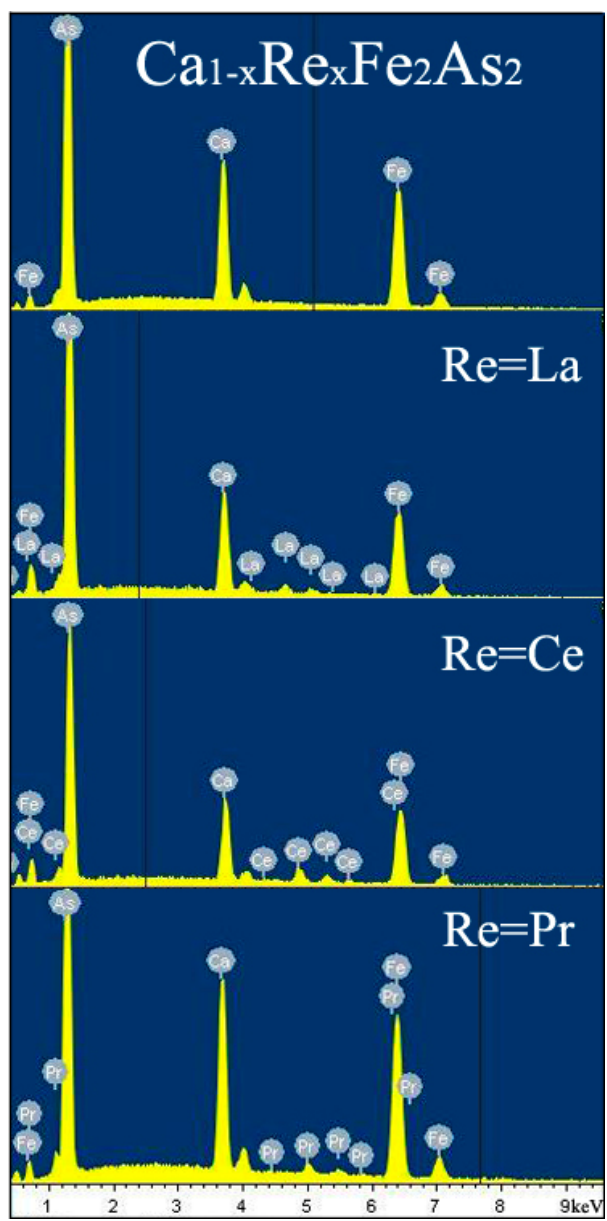

Fig.2 Qi et al.

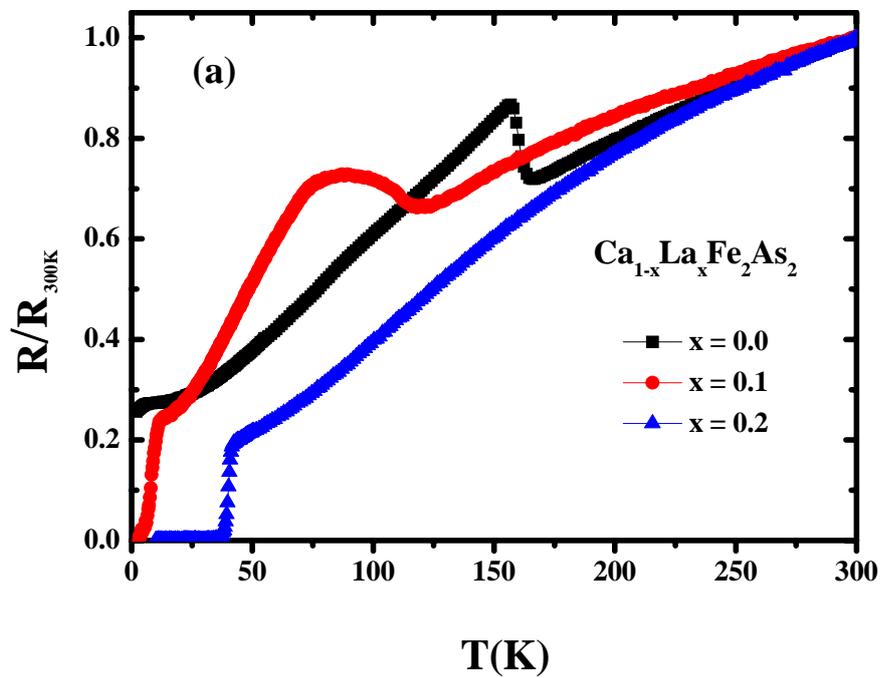
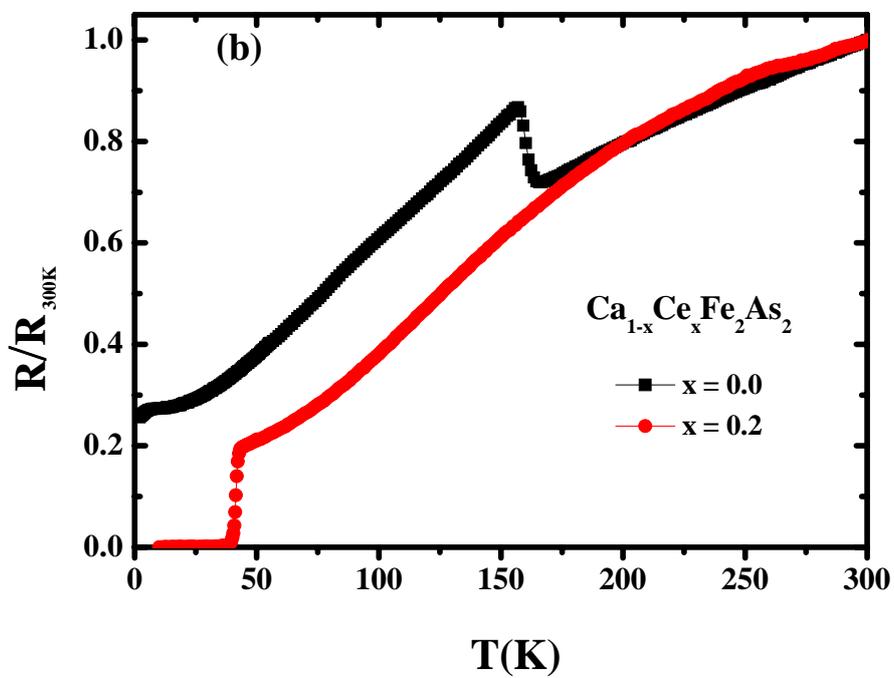

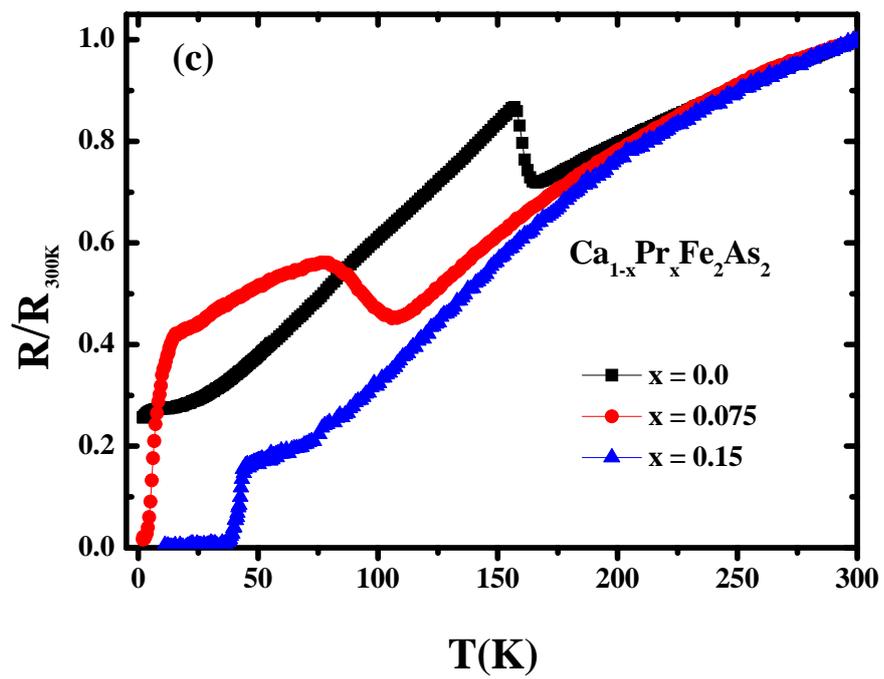

Fig.3 Qi et al.

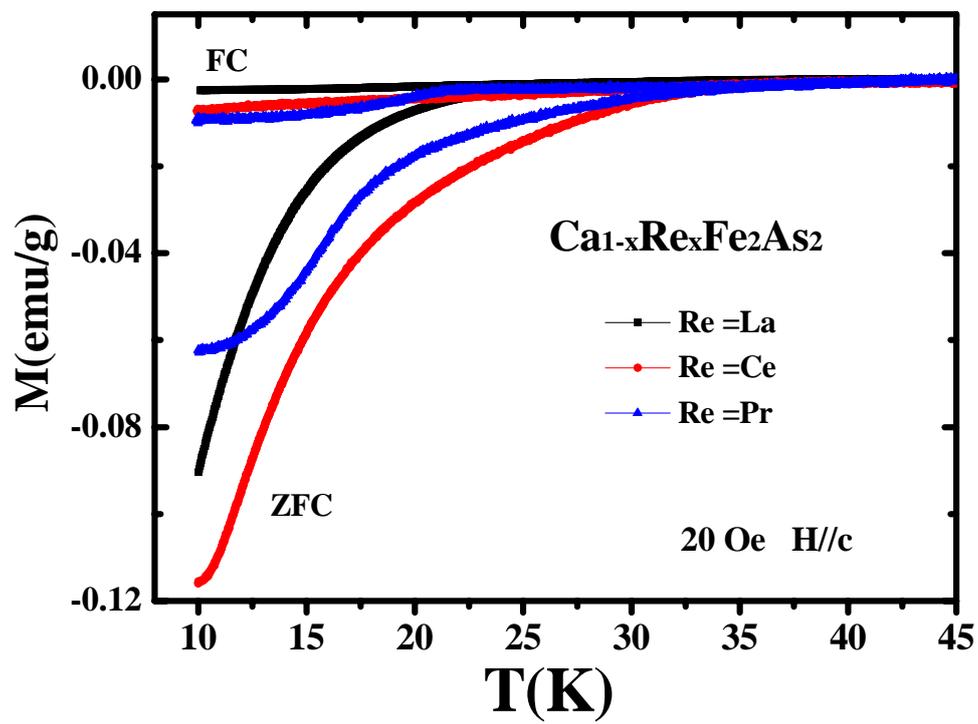

Fig.4 Qi et al.

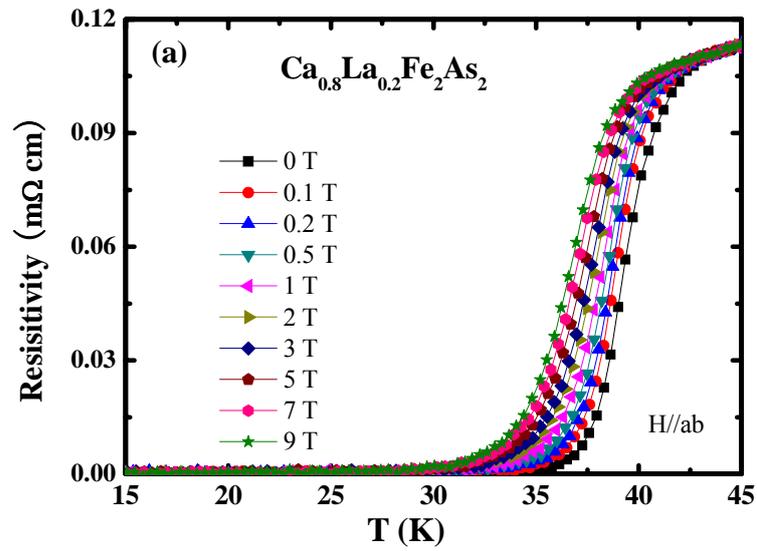
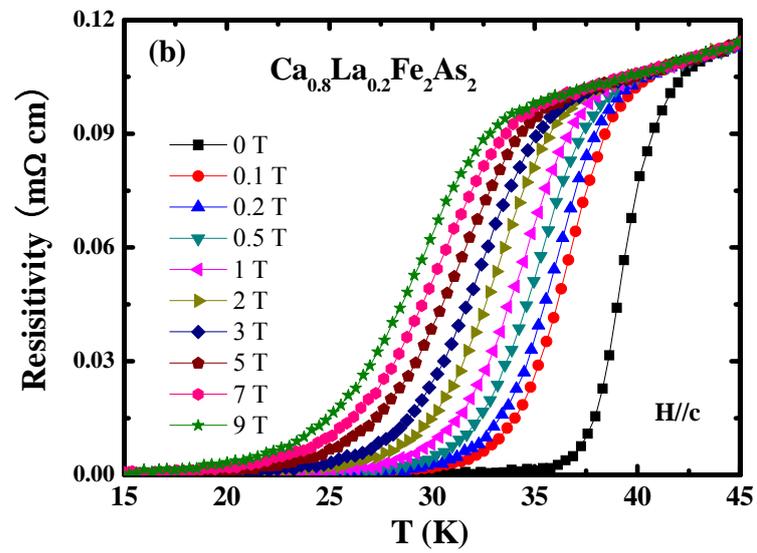

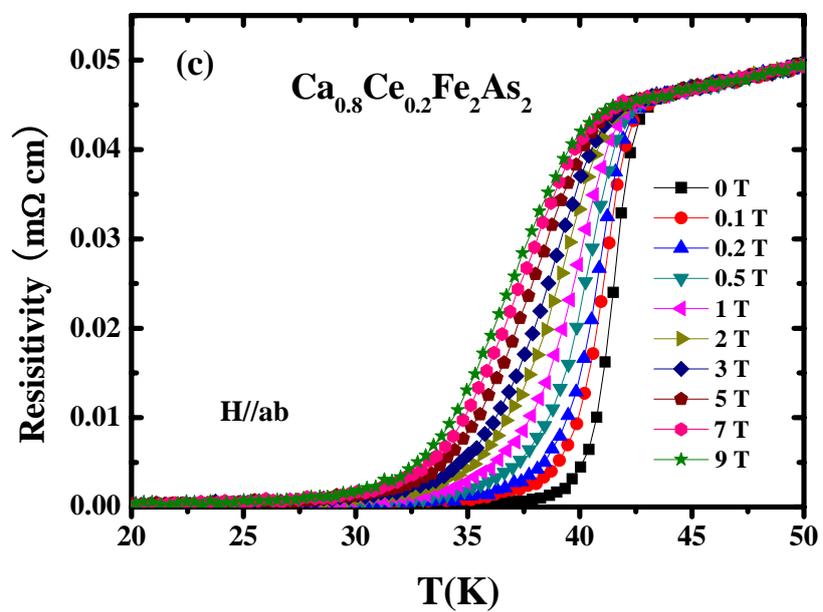

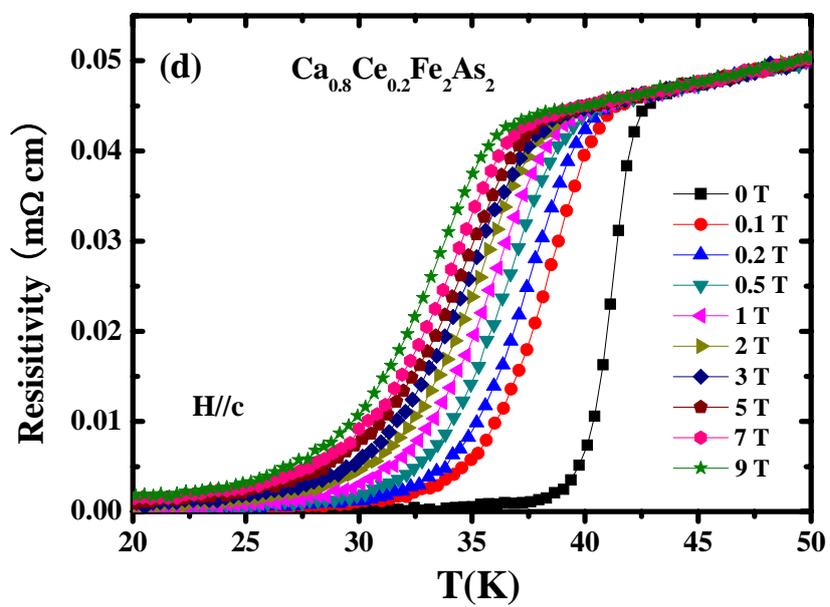

Fig.5 Qi et al.

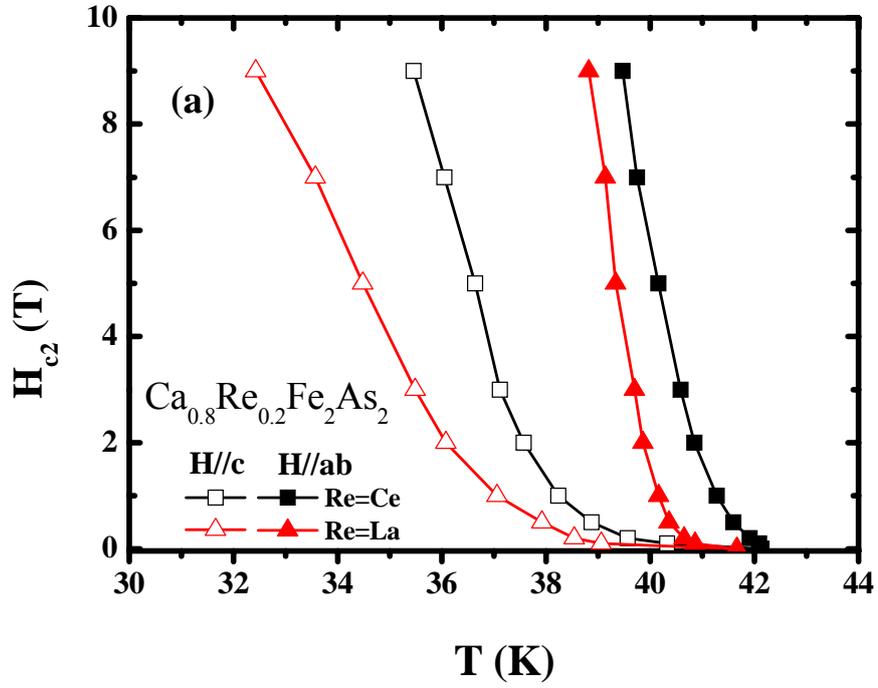

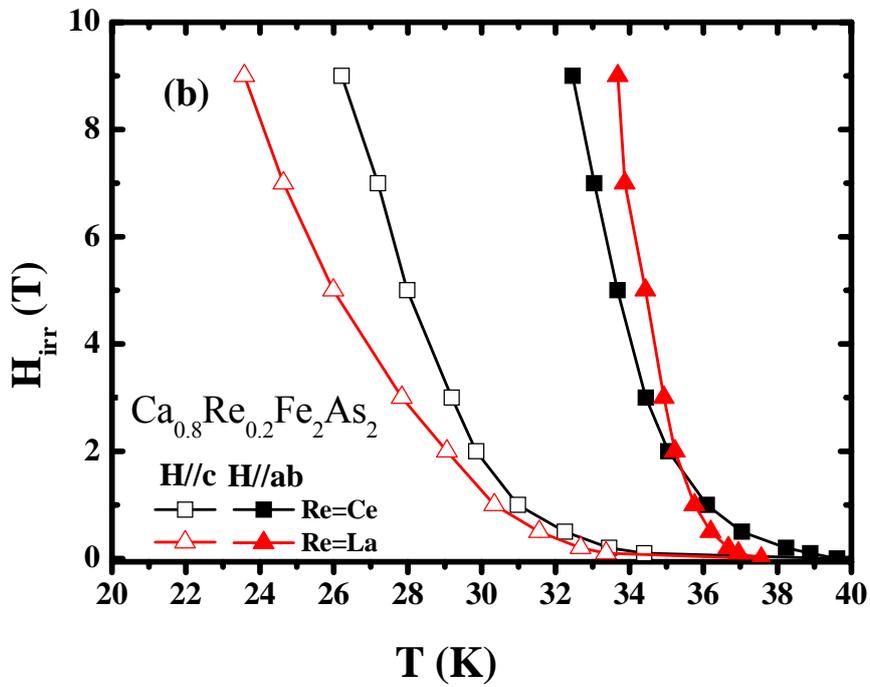

Fig.6 Qi et al.

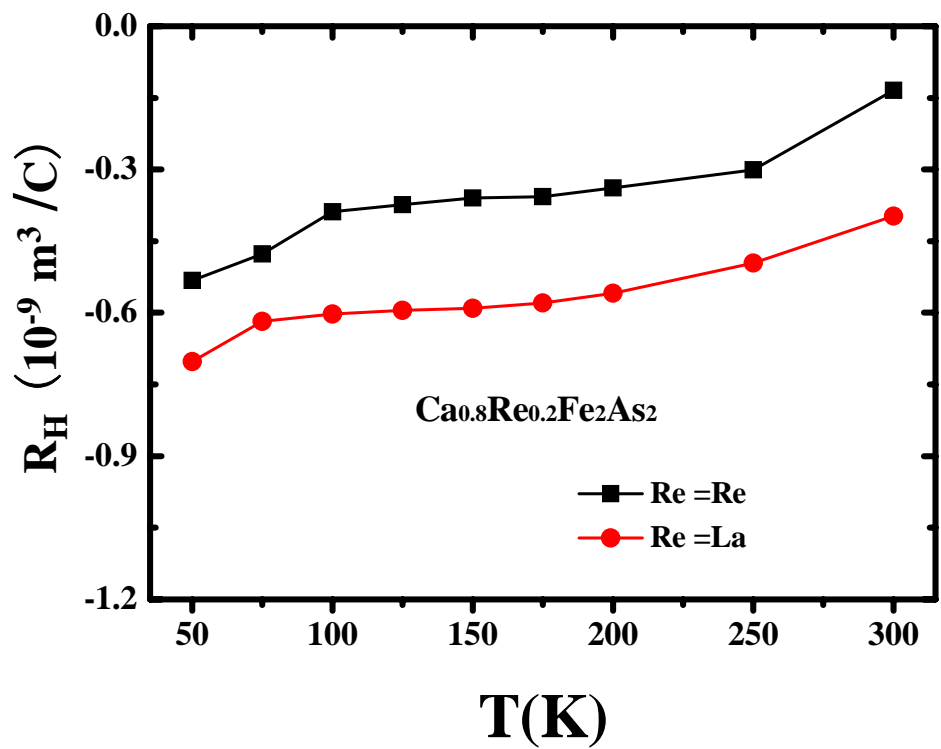

Fig.7 Qi et al.